\documentclass{PoS}
\usepackage{braket}
\usepackage{float}
\usepackage{graphicx}
\usepackage{bm}
\usepackage{amssymb}
\usepackage{amsmath}
\usepackage[numbers,sort&compress]{natbib}
\title{Extracting Neutron Polarizabilities from Compton Scattering on The Quasi-Free Neutrons in $\gamma \text{d}\rightarrow\gamma \text{n p}$}

\ShortTitle{Extracting Neutron Polarizabilities from Compton Scattering on The Quasi-Free Neutrons in $\gamma \text{d}\rightarrow\gamma \text{n p}$}

\author{\speaker{Berhan Demissie}\\
         \\Institute for Nuclear Studies, Department of Physics, George
  Washington University, Washington DC 20052, USA;
        E-mail: \email{berhan@gwmail.gwu.edu}}

\author{Harald W.~Grie{\ss}hammer%\thanks{}
  \\Institute for Nuclear Studies, Department of Physics, George
  Washington University, Washington DC 20052, USA;
        E-mail: \email{hgrie@gwu.edu}}

\abstract{Compton scattering processes are ideal to study electric and magnetic dipole polarizabilities of nucleons. These fundamental quantities parametrize the responses of the internal degrees of freedoms to a monochromatic photon probe. In this work, the inelastic channel  $\gamma d\rightarrow \gamma np$ is treated in $\chi$EFT, with a focus on the kinematic region of the neutron quasi-free peak, where the momentum of the outgoing proton is small enough that it is considered to remain at rest. This provides access to the Compton scattering process $\gamma n \rightarrow \gamma n$, from which the scalar polarizabilites of the neutron, $\alpha_{E1}$ and $\beta_{M1}$, are extracted. Using $\chi$EFT, triple differential cross-sections $d^3\sigma/dE_nd\Omega_{\gamma'} \Omega_n$ in the photon energy range of 200-to-400 MeV are computed. The biggest contribution comes from the impulse approximation, with small corrections stemming from final state interactions, meson exchange currents and rescattering. In this presentation, we report progress and preliminary results. }

\FullConference{The 8th  International Workshop on Chiral Dynamics \\
                 29 June 2015 - 03 July 2015\\
                 Pisa, Italy}

\begin{document}

\section{Introduction and Background}

Nucleon polarizabilities are fundamental quantities, much like mass and charge, that parameterize the responses of the charge and current distribution as well as of the spin to an external electromagnetic field. In this work, we focus on the electric polarizability, $\alpha_{E1}$, and the magnetic polarizability, $\beta_{M1}$, collectively called scalar polarizabilities. Various nuclear and cosmological processes rely on accurate values of these quantities for their proper description. Isospin breaking in the nucleon mass difference $M_n-M_p$~\cite{walker_loud_nuclear_2014,walker-loud_electromagnetic_2012,Gasser_cottingham_2015}, the proton radius puzzle~\cite{pohl_size_2010}, and the  magnetization of neutron stars~\cite{Bernabeu_intrinsic_1974} are but a few examples whose theoretical treatment was shown to be limited by the accuracy to which we know the scalar polarizabilites.   

Nuclear Compton scattering processes are ideal to explore the internal structure of a nucleon. The long-wavelength amplitude of a $\gamma N \rightarrow \gamma N$ interaction is dominated by the Thomson term which depends only on the charge and mass of the nucleon. As the energy of the probing photon increases, the low energy internal degrees of freedoms~(d.o.f.s) that make up the charge and current distribution as well as the spin, become more prominent. 

When the photon interacts with a nucleon, the electric field rearranges the charge distribution inside the nucleon such that the positive charges separate from the negative ones along the direction of the field. This induces an electric dipole moment. Similarly, the magnetic field of the photon rearranges the current distribution, inducing a magnetic dipole moment. The induced dipoles are proportional to the external fields:
\begin{equation}
\vec{d}(\omega) = \alpha_{E1}(\omega) \vec{E}(\omega) \hspace{2cm} \vec{\mu}(\omega) = \beta_{M1}(\omega) \vec{B}(\omega)
\label{dip}
\end{equation}
The proportionality ``constants" are the energy-dependent scalar polarizabilites. They parameterize how stiff the charge and current distributions are against the distorting effects of the external fields at a fixed frequency. 

The induced electric or magnetic dipoles, in turn, re-radiate at the same frequency with a signature angular dependence unique to the dipole. Similarly, the dipole deformations in the spin d.o.f.s by the external electric and magnetic fields are parametrized by spin polarizabilites.

The characteristic angular dependencies associated with transitions of photon multipolarities ${Xl\to Yl^\prime}$ (where $X,Y = E,M \text{ and } l^\prime = {l \pm\{0,1\}} $) allow for the interpretation and identification of nucleon polarizabilites. The multipole analysis of Compton scattering is related to the structure parts of the Compton amplitudes~\cite{grieshammer_using_2012,hildebrandt_signatures_2004}. For detailed discussion, readers are advised to refer to the proceedings of the talk by Grie{\ss}hammer~\cite{haralpos} and of the plenary talk given by McGovern~\cite{mcgovpos}.

In the past two decades, chiral Effective Field Theory~($\chi$EFT), the low energy approximation of QCD that includes baryons, has become a robust theory for the treatment of the interaction of nucleons with photons. The upper limit of the low energy regime is set by the chiral symmetry breaking scale, $\Lambda_{\chi} \sim$ 1GeV. $\chi$EFT does not describe the details of any physics beyond its range of applicability, but it does parametrize the high energy regime through low energy constants (LECs) that are fixed by first principle or data. The ratio of low (small momentum transfer, $p$; pion mass, $m_\pi$; $\Delta$-N mass splitting, $M_\Delta -M_N$) and high ($\Lambda_{\chi}$) momentum components of the theory form a small expansion parameter by which the interaction Lagrangian is ordered perturbatively. The Lagrangian itself is governed by the symmetries of the underlying theory, namely QCD. These basic attributes make $\chi$EFT a model-independent theory with well defined uncertainties~\cite{bernard_chiral_1995,grieshammer_using_2012}. 
In what is considered an early success of $\chi$EFT, it predicted the proton polarizabilites at first order and with no unknown LECs. These results were in good agreement with experimental findings~\cite{bernard_chiral_1995}.

Currently, the elastic Compton scattering on the proton is computed to $\mathcal{O}(e^2\delta^4)$ where $\delta$ is the small expansion parameter ($\delta \equiv \frac{M_\Delta-M_N}{\Lambda_{\chi}} = (\frac{m_\pi}{\Lambda_\chi})^{1/2}$), with the $\Delta$(1232) resonance included as dynamic degrees of freedom~\cite{mcgovern_compton_2013}. By fitting the theory to the wealth of proton data from Compton scattering experiments on a hydrogen target, the following values for the proton polarizabilities are extracted (in the canonical units of $10^{-4}\text{fm}^3$)~\cite{mcgovern_compton_2013}:
\begin{eqnarray}
\alpha_p = 10.65 \pm 0.35({\text{stat}}) \pm 0.2({\text{Baldin}})\pm 0.3({\text{theory}}) \nonumber
\\
\beta_p = 3.15 \mp 0.35({\text{stat}}) \pm 0.2({\text{Baldin}})\mp 0.3({\text{theory}}) 
\label{eqn_0}
\end{eqnarray}

The neutron values, on the other hand, are not that easy to measure, largely due to the lack of a stable neutron target. The neutron polarizabilites are extracted by using elastic and inelastic Compton scattering on light nuclei, such as the deuteron and ${}^3$He. In $\gamma d \rightarrow \gamma' d$, the isoscalar polarizabilites, $\alpha_s =\frac{1}{2} (\alpha_p+\alpha_n)$ and $\beta_s = \frac{1}{2}(\beta_p + \beta_n)$, of the bound system are extracted from the cross-section. By subtracting off the known values for the proton, the neutron polarizabilites are obtained. The polarizabilites of the neutron as determined from elastic Compton scattering on the deuteron using $\chi$EFT, are~\cite{Myers_measurement_2014}:
\begin{eqnarray}
\alpha_n = 11.55 \pm 1.25({\text{stat}}) \pm 0.2({\text{Baldin}})\pm 0.8({\text{theory}}) \nonumber
\\
\beta_n = 3.65 \mp 1.25({\text{stat}}) \pm 0.2({\text{Baldin}})\mp 0.8({\text{theory}}) 
\label{eqn_00}
\end{eqnarray}

In the following, the $\chi$EFT treatment of the inelastic Compton scattering on the deuteron is introduced.

\section{Neutron Quasi-Free Peak}
Extracting neutron polarizabilities from elastic Compton scattering on the deuteron involves disentangling the proton and nuclear binding contributions to the amplitude and subtracting them off. On the other hand, when considering the inelastic Compton scattering reaction, $\gamma + d \rightarrow \gamma' + n + p $, like we do in this project, the approach is to minimize, if not eliminate, the contributions of the proton and of nuclear binding from the amplitude altogether so that neutron interactions are measured more directly.

This is achieved in the kinematic region where the proton momentum $\vec{p}_p \rightarrow 0$. The center of this region, $\vec{p}_p =0$, is called the neutron quasi-free peak (nqfp), in which the incoming photon knocks the neutron out while the proton remains at rest. This impulse approximation (IA) effectively reduces the inelastic Compton process to a weighted elastic Compton scattering on the neutron, whose amplitude is given symbolically by: 
\begin{equation}
\braket{\gamma n| T| \gamma' n} \Psi({\vec{p}_p})
\label{eq_2}
\end{equation}
where $\Psi({\vec{p}_p})$ is the deuteron wave function at the momentum of the spectator proton.

Levchuk et al.~\cite{levchuk_photon_1994} outlined the kinematic constraint as well as the photon energy range, 200-to-400 MeV, where sensitivity to neutron polarizabilites is expected to be high. A quasi-free experiment was conducted at SAL by Kolb et al. that served as a feasibility study (one data point)~\cite{sal}, followed by Kossert et al. at MAMI~\cite{kossert_quasi-free_2003}, in the kinematics recommended by Ref.~\cite{levchuk_photon_1994}. Neutron polarizabilites were extracted from the latter experiment, which is the most definitive quasi-free experiment to date (9 data points).

However, there are significant model assumptions in the dispersion-relation calculations used for the single nucleon amplitudes of this analysis. Furthermore, the model uncertainties reported in~\cite{kossert_quasi-free_2003} are believed to be underestimated~\cite{grieshammer_using_2012}. 

The purpose of our project is to re-analyze the inelastic Compton scattering on the deuteron in the nqfp region using the heavy baryon version of $\chi$EFT. The model independence of $\chi$EFT together with its well-defined reproducible theoretical uncertainties makes it the ideal platform for this purpose.

\begin{figure}[H]
 \centering
 \includegraphics[width=0.5\textwidth]{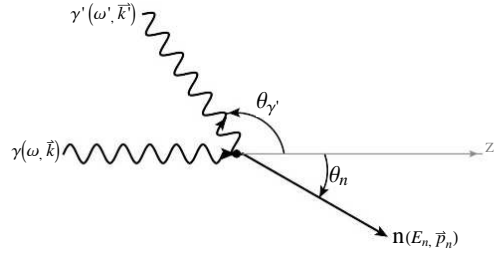}
 \caption{Schematic representation of the $\gamma d \rightarrow \gamma n d$ at the nqfp kinematic region.}
\label{fig3}
\end{figure}

In the nqfp region, the spectator proton simplifies the kinematics such that the scattering process takes place in a plane. The reaction is demonstrated in Fig.~\ref{fig3}. The total energy and three-momentum of each participating particle is labeled in the figure. The scattering angle of the outgoing photon $\theta_{\gamma'}$ and that of the recoiling neutron $\theta_n$ are also shown. Unless stated otherwise, all schematic and kinematic variables are given in the lab frame. 

\begin{figure}[H]
 \centering
 \includegraphics[width=0.3\textwidth]{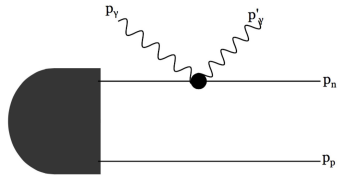}
\caption{IA: the incoming photon interacts only with one of the nucleon, in the nqfp case, knocking the neutron out while the proton remains a spectator.}
\label{fig_1}
\end{figure}

The dominating contribution to the amplitude is the IA as depicted in Fig.~\ref{fig_1}. 

\begin{figure}[H]
 \centering
 \includegraphics[width=0.9\textwidth]{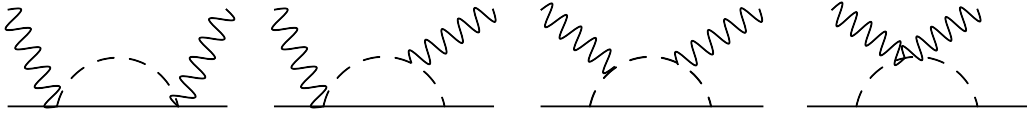}
\caption{Third order $\pi$N loop graphs contributing at $\mathcal{O}(e^2\delta^1)$. Crossed diagrams not shown.}
\label{fig_2}
\end{figure}

\begin{figure}[H]
 \centering
 \includegraphics[width=0.95\textwidth]{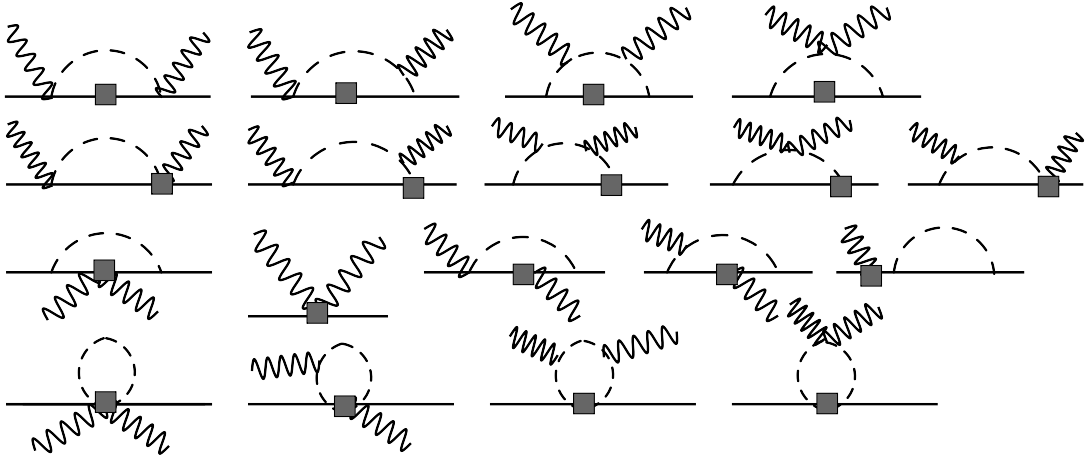}
\caption{Fourth order $\pi$N loop graphs $\mathcal{O}(e^2\delta^2)$. Crossed diagrams not shown.}
\label{del}
\end{figure}

\begin{figure}[H]
 \centering
 \includegraphics[width=0.3\textwidth]{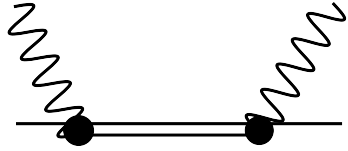}
\caption{s-channel $\Delta(1232)$-pole graph. The blobs: $\gamma$N$\Delta$ couplings including vertex corrections~\cite{mcgovern_compton_2013}.}
\label{pole}
\end{figure}

In addition to the nucleon pole diagrams, the leading order structure contribution to $\braket{\gamma n| T| \gamma' n}$ are the $\pi$N-loops at $\mathcal{O}(e^2\delta^2)$ shown in Fig.~\ref{fig_2} and~\ref{del}. In the energy range considered, the effect of the $\Delta(1232)$ resonance is dominant. Thus, it is included as an explicit degree of freedom with non-zero width, following Ref.~\cite{mcgovern_compton_2013}. Here, the $\Delta(1232)$ propagates close to its mass shell. Therefore, the Delta pole graph, Fig.~\ref{pole}, is included at the order $\mathcal{O}(e^2\delta^{-1})$ with the propagator resummed giving the $\Delta(1232)$ its non-zero width. 

\begin{figure}[H]
 \centering
 \includegraphics[width=0.9\textwidth]{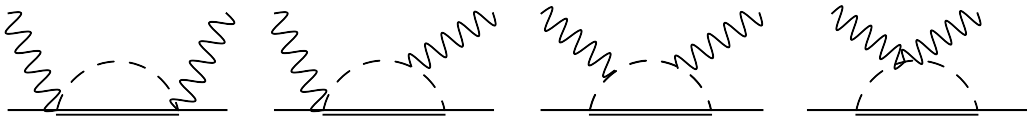}
\caption{$\pi\Delta$-loop graphs contributing at $\mathcal{O}(e^2\delta^1)$.}
\label{fig_3}
\end{figure}

Furthermore, the $\pi \Delta$-loop graphs, shown in Fig.~\ref{fig_3}, contribute at $\mathcal{O}(e^2\delta^1)$. For an in-depth discussion, readers are referred to the proceeding from the plenary talk given by J. McGovern~\cite{mcgovpos} and Ref.~\cite{mcgovern_compton_2013}.

\begin{figure}[H]
 \centering
 \includegraphics[width=0.7\textwidth]{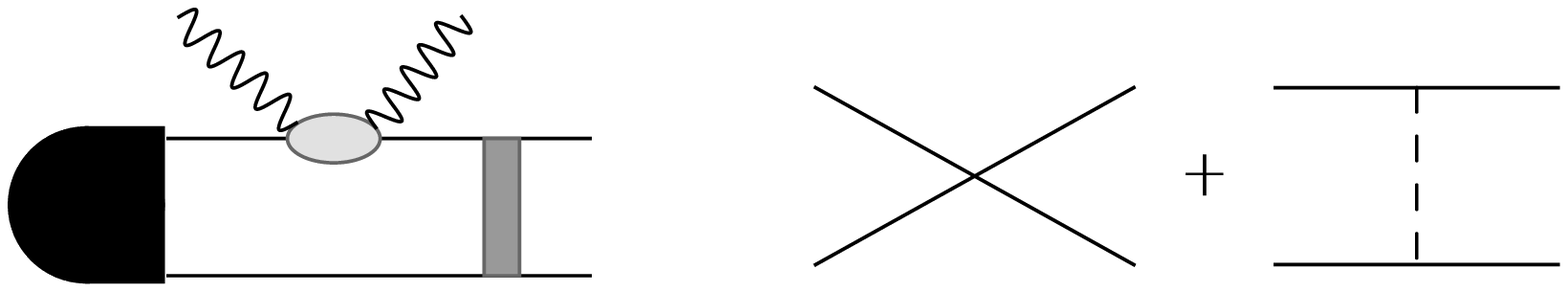}
\caption{(a) final state interaction  \hspace{3cm}(b) LO NN-interaction}
\label{fig_4}
\end{figure}

Final state interactions~(FSIs) and meson exchange currents (MECs) are corrections to the IA amplitudes. FSIs are those where $\gamma N \rightarrow \gamma N$ is followed by the interaction of outgoing nucleons with each other before they fly out. The interactions between the nucleons are dominated by the leading order NN-interaction; see Fig.~\ref{fig_4}. The contribution of final state interactions are expected (and are shown later) to be small. When $\vec{p}_p = 0$ and in the energy range considered, the outgoing neutron carries most of the energy and momentum transfer. It leaves the process so fast that there is no time for complex NN-interaction to take place. 

As motivated above, only the 1$\pi$ exhange and the four-nucleon vertex term are included for the NN-interaction, Fig.~\ref{fig_4}b. The vertext is given by a contact term $C_0 = C_s + \sigma_1 \cdot \sigma_2C_v$. The LECs, $C_s$ and $C_v$ are fixed by using the np~bound system (deuteron) and the np~scattering length.

\begin{figure}[H]
 \centering
 \includegraphics[width=0.9\textwidth]{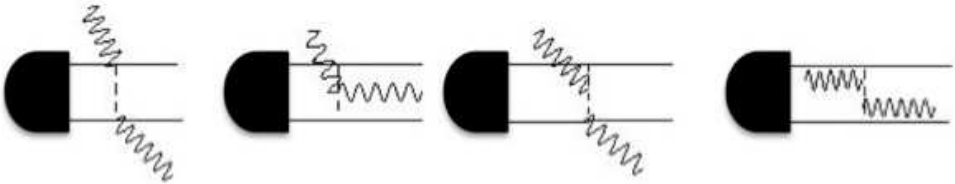}
\caption{Meson exchange current (MEC) diagrams. Crossed diagrams not shown.}
\label{fig_5}
\end{figure}

Following the same argument and the analysis of the contributions of FSIs and of MECs, discussed in the following section, rescattering diagrams are expected to be negligible, and are thus not included in our calculation. In MEC corrections, the two nucleons interact through a pion exchange. At the order we are working, the relevant MECs are shown in Fig.~\ref{fig_5}.  

\section{Preliminary Results}
The IA, and FSI as well as  MEC corrections were used to calculate the triple differential cross-section for the incoming photon energy ranges from 200-to-400 MeV and the scattering angle $\theta_\gamma = 136^0$.
In the following, our preliminary analysis of the triple differential cross-section $ \frac{d^3\sigma}{dE_nd\Omega_nd\Omega_{\gamma'}}$ is presented, with $\alpha_n$ and $\beta_n$ fixed to the values in Eq.~\eqref{eqn_00}.

\begin{figure}[H]
 \centering
 \includegraphics[width=0.5\textwidth,trim={0 0 0 0},clip]{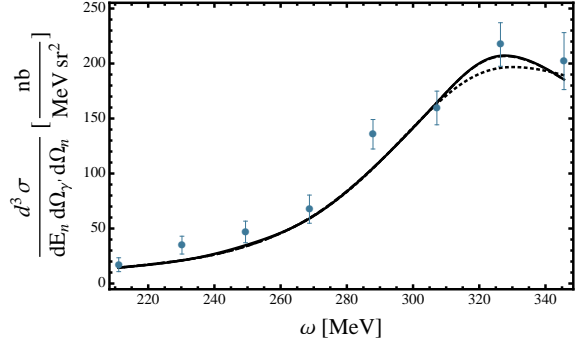}
\caption{The effect of the contributing amplitudes on the nqfp triple differential cross-section of the quasi-free neutron. Dotted: IA; dashed: FSI added; solid: combined contribution of IA, FSI and MEC. }
\label{fig_6}
\end{figure}

The dotted line in Fig.~\ref{fig_6} shows the cross-section for the impulse approximation. The effect of the FSI correction onto the IA is the barely visible dashed curve which essentially overlaps with the IA curve. The solid curve corresponds to the cross-section with IA, FSI and MEC put together. As expected, we see that the addition of FSI and MEC to the IA amplitude has a small effect on the cross-section. This supports the choice to describe the NN-interaction by the LO only and validates the decision to not include higher order rescattering interactions. The cross-section for which all amplitudes are included describes the data with a $\chi^2$ per d.o.f of about 1.2. 

A similar kinematic constraint, $\vec{p}_n = 0 $, can be put on the process $\gamma + d \rightarrow \gamma' + n + p $ in which, instead of the proton, the neutron becomes the spectator and the incoming photon interacts predominantly, if not entirely, with the proton. In a similar vein, the region is called proton quasi-free peak (pqfp) and the FSI and MEC contributions become corrections to the IA amplitudes on the proton. 

\begin{figure}[H]
 \centering
 \includegraphics[width=0.5\textwidth]{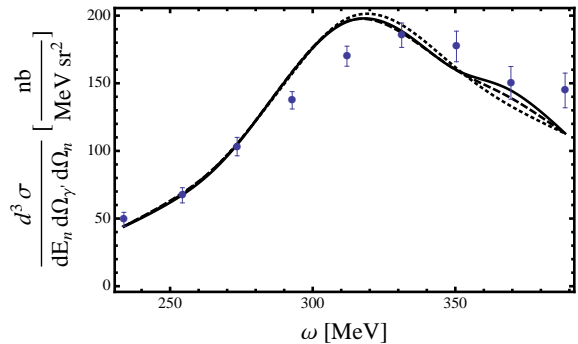}
\caption{The effect of the contributing amplitudes on the pqfp triple differential cross-section. Dotted: IA; dashed: FSI added; solid: combined contribution of IA, FSI and MEC.}
\label{fig_7}
\end{figure}

\begin{figure}[H]
 \centering
 \includegraphics[width=0.5\textwidth]{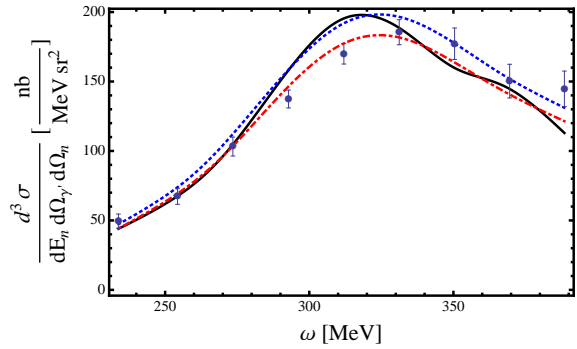}
\caption{Triple differential cross-section for pqfp. Solid, black line: $\chi$EFT; dashed, blue: model from Ref.~\cite{levchuk_photon_1994}; dot-dash, red: model from Ref.~\cite{levchuk_photon_1994}, scaled by a factor of 0.93.}
\label{fig_8}
\end{figure}

Here again, we computed the cross-sections for the inelastic process in the pqfp region. This time $\alpha_p \text{ and } \beta_p$ were fixed to the values in Eq.~\eqref{eqn_0}. The preliminary result is shown in Fig.~\ref{fig_7}. The points are quasi-free proton data from Kossert et al.~\cite{kossert_quasi-free_2003}. There is discrepancy between data and theory. The $\chi^2$ per d.o.f is about 3.4. However, the authors saw a similar discrepancy between the model of Ref.~\cite{levchuk_photon_1994} and the data. Fig.~\ref{fig_8} compares our $\chi$EFT curve (solid) with that of Ref.~\cite{levchuk_photon_1994}~(dotted). Both curves show similar deviation from data, in particular where the most deviations occur (290-340 MeV). In Ref.~\cite{kossert_quasi-free_2003}, the authors note that if the model were to be scaled by 0.93, the scaled curve (dot-dashed) would better describe the data. We have so far not found a compelling explanation for the scaling factor. 

 We have yet to explain why the proton data deviates from our description.  

At this point our analysis is preliminary and we do not make conclusive remarks. However, going forward, we will perform a two-parameter fit and a one-parameter fit with the Baldin sum rule as constraint in order to extract neutron polarizabilites, $\alpha^n_{E1} \text{ and }\beta^n_{M1}$, as well as $\gamma^n_\pi$ $-$ the backward spin polarizability of the neutron. The pqfp data will also be fitted for proton values. By comparing neutron and proton results, we will also learn more about the discrepancy in the proton analysis.

\acknowledgments
The speaker thanks the organizers for making the conference a cultural and intellectual experience. This work is supported by the US Department of Energy under contracts DE-FG02-95ER-40907, and by the Dean's Research Chair programme of the Columbian College of Arts and Sciences of The George Washington University.

%\bibliographystyle{plainmod,unsrt}
%\bibliography{bdemissielst}
\end{document}